%% file: MICCAI2025_paper_template.tex
\documentclass[runningheads]{llncs}

\usepackage{amsmath}
\usepackage{booktabs}
\usepackage[T1]{fontenc}
\usepackage{multirow}
\usepackage{colortbl}
\usepackage{graphicx,verbatim}
\usepackage{microtype}
\usepackage[table]{xcolor} 
\definecolor{headercolor}{gray}{0.92}
\definecolor{yellowcolor}{HTML}{FFF9DB} 

\newcommand{\ourmodel}{PRETI}

\begin{document}
\title{PRETI: Patient-Aware Retinal Foundation Model via Metadata-Guided Representation Learning}
\titlerunning{PRETI: Patient-Aware Retinal Foundation Model via Metadata}

\author{
Yeonkyung Lee\inst{1} \and
Woojung Han\inst{1} \and
Youngjun Jun\inst{1} \and \\ 
Hyeonmin Kim\inst{2} \and
Jungkyung Cho\inst{2} \and
Seong Jae Hwang\inst{1}
}

\authorrunning{Lee et al.}  

\institute{
Yonsei University, Seoul, South Korea \\
\email{\{yeonkyung.lee, dnwjddl, youngjun, seongjae\}@yonsei.ac.kr}
\and
Mediwhale, Seoul, South Korea \\
\email{\{luke.kim, kyla.cho\}@mediwhale.com}
}

\maketitle              
\input{sec/0_abstract}
\input{sec/1_into}
\input{sec/2_method}
\input{sec/3_experiment}
\input{sec/4_conclusion}

\clearpage
\bibliographystyle{splncs04}
\bibliography{mybibliography}
\end{document}

%% file: sec/0_abstract.tex
\begin{abstract}
Retinal foundation models have significantly advanced retinal image analysis by leveraging self-supervised learning to reduce dependence on labeled data while achieving strong generalization. Many recent approaches enhance retinal image understanding using report supervision, but obtaining clinical reports is often costly and challenging. In contrast, metadata (e.g., age, gender) is widely available and serves as a valuable resource for analyzing disease progression. To effectively incorporate patient-specific information, we propose \ourmodel{}, a retinal foundation model that integrates metadata-aware learning with robust self-supervised representation learning. We introduce Learnable Metadata Embedding (LME), which dynamically refines metadata representations. Additionally, we construct patient-level data pairs, associating images from the same individual to improve robustness against non-clinical variations. To further optimize retinal image representation, we propose Retina-Aware Adaptive Masking (RAAM), a strategy that selectively applies masking within the retinal region and dynamically adjusts the masking ratio during training. \ourmodel{} captures both global structures and fine-grained pathological details, resulting in superior diagnostic performance. 
Extensive experiments demonstrate that \ourmodel{} achieves state-of-the-art results across diverse diseases and biomarker predictions using in-house and public data, indicating the importance of metadata-guided foundation models in retinal disease analysis. Our code and pretrained model are available at https://github.com/MICV-yonsei/PRETI

\keywords{Foundation Model  \and Masked Image Modeling  \and Metadata}
\end{abstract}

%% file: sec/1_into.tex
\section{Introduction}
Foundation models have recently gained attention for their remarkable adaptability across various tasks~\cite{dino,llava,clip}.
By training on extensive unlabeled data, they capture core patterns and relationships that can be fine-tuned with minimal effort for various downstream applications~\cite{qlora,llama-adapter}.
While their applications in areas like computer vision and natural language processing have shown promising results~\cite{dino,sam,clip}, their use in medical contexts remains relatively unexplored.
 
A key motivation for developing medical foundation models lies in the scarcity of annotated medical data, as labeling clinical images requires expert knowledge and is both time-consuming and expensive. Recently, medical foundation models have shown the potential of using large-scale unlabeled data \cite{GMAI,retfound}. 
Especially in ophthalmology, retinal foundation models are being studied due to the discovery that clinical features associated with ocular diseases can be observed in color fundus photography (CFP)~\cite{eye_is_anything2,eye_is_anything1}.
RETFound~\cite{retfound} demonstrated how image-based pre-training can serve as a strong foundation for retinal fundus analysis.
Expanding on this, foundation models such as FLAIR~\cite{flair}, RETCLIP~\cite{retclip}, and UrFound~\cite{urfound} incorporate report data, enhancing the multimodal understanding of retinal diseases.
While these approaches have significantly advanced retinal image analysis through report supervision, expert descriptions of patients such as clinical diagnostic reports are difficult to obtain, thereby reducing the amount of available data.
On the other hand, among various types of metadata, age and gender serve as minimal patient information that can be easily obtained. Moreover, they exert a critical influence on understanding how ocular pathologies develop, progress, and respond to treatment \cite{kyla_1,kyla_2}.
Thus, we incorporate metadata into foundation models to enable them to capture patient-specific variations.

Despite the significance of metadata, research on foundation models leveraging metadata has been relatively scarce \cite{metadata_embedding,metadata_contrastive1,metadata_contrastive2}. 
These approaches treat metadata as static embeddings~\cite{metadata_embedding} or incorporate it through a contrastive loss criterion~\cite{metadata_contrastive1,metadata_contrastive2}. 
However, relying on metadata in such a static manner fails to fully leverage metadata for capturing complex image-metadata relationships.
Instead, we introduce the \textit{Learnable Metadata Embedding (LME)}. LME dynamically integrates metadata into the learning process, allowing it to adapt and evolve alongside visual features.
In contrast to static embeddings or fixed constraints, LME enables a more flexible representation of retinal data. Specifically, we focus on the two key metadata variables (age and gender) which are fundamental to understanding retinal disease patterns~\cite{kyla_metadata_1,kyla_metadata_2}.

Enhancing representation learning through metadata is a notable step forward. However, retinal foundation models still face challenges from imaging variations due to device differences, lighting, and inter-eye anatomical distinctions.
To address these issues, we construct \textit{patient-level data pairs}, grouping images from the same individual, even when captured with different cameras. Additionally, considering the clinical symmetry between the left and right eyes in ocular diseases \cite{Glaucoma-symmetry,DR-symmetry,AMD-symmetry}, we treat them as equivalent when forming patient-level pairs. 
This strategy accounts for non-disease-related variations, enabling the model to learn robust representations.
To effectively utilize both metadata and patient-level pairs, we build upon a self-supervised modeling framework based on Siamese Masked Autoencoder (SiamMAE) \cite{siamese_mae}, an extension of MAE~\cite{mae}. 
This architecture suits our patient-level pairs by disrupting one image and restoring it using information from the other, learning visual correspondence between both eyes.

While MAE is a self-supervised learning method that learns image features by reconstructing images from randomly masked inputs, its standard random masking strategy does not align well with retinal images, where important structures are confined to the eye region. To address this, we implement \textit{Retina-Aware Adaptive Masking (RAAM)}, a technique designed to align with the specific properties of retinal image data. 
RAAM selectively applies masking within the retinal region and dynamically adjusts the masking ratio during training. This refinement enhances the ability of the model to capture both global structures and fine details, making it better suited for retinal image analysis.

Based on patient-level pairs incorporating LME and RAAM, we confidently propose \ourmodel{}, a foundation model that focuses on CFP and metadata for comprehensive assessment. Through extensive experiments, we validate its state-of-the-art performance across diverse disease and biomarker prediction tasks.

%% file: sec/2_method.tex
\section{Methods}
\begin{figure}[t]
  \centering
  \includegraphics[width=\linewidth]{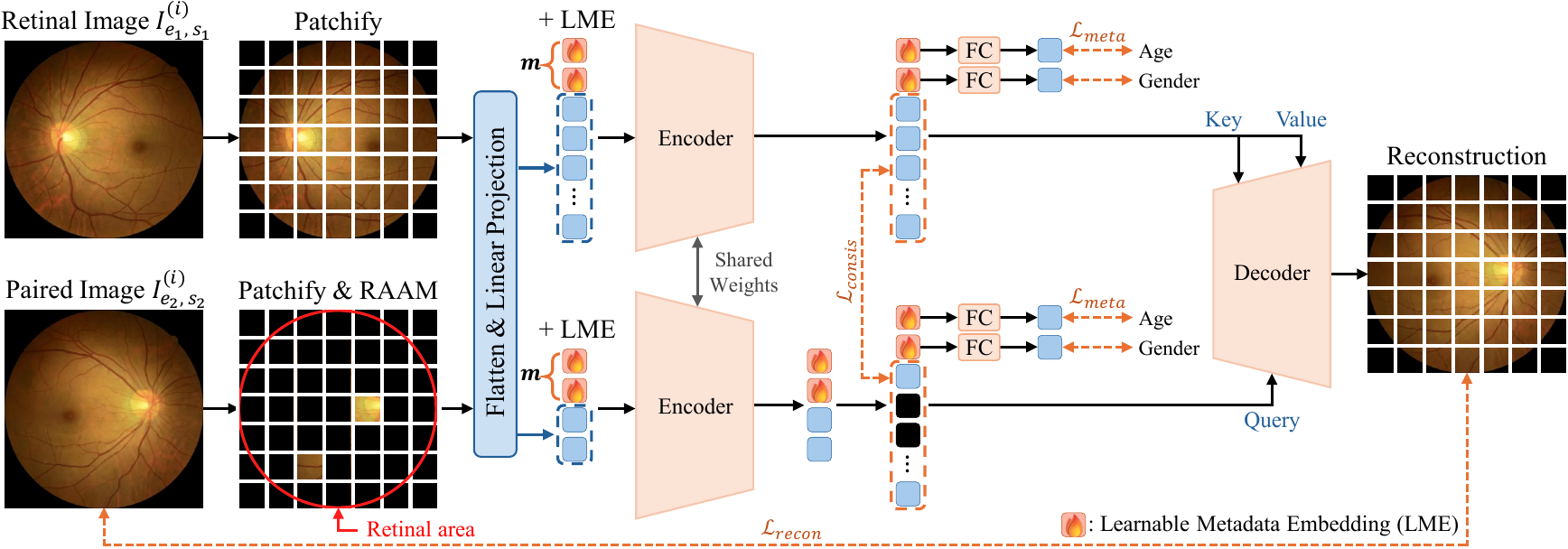}
  \caption{
  The architecture of \ourmodel{}, a foundation model for retinal image representation learning, illustrates the integration of color fundus photography (CFP) and metadata through Learnable Metadata Embedding (LME). The model processes paired retinal images using Patient-level Pairs and employs shared encoders with dynamic metadata adaptation (age and gender). It also incorporates the Retina-Aware Adaptive Masking (RAAM) strategy for selective masking and reconstructing images via a decoder to enhance representation learning for robust clinical generalization.
  }
  \label{fig:arch}
\end{figure}

Our foundation model, \ourmodel{}, integrates visual and metadata information for retinal image representation learning (Fig.~\ref{fig:arch}). We first introduce patient-level data pairs, followed by the retina-aware adaptive masking strategy (RAAM), which is designed specifically for retinal images. Then, we present the learnable metadata embedding (LME) for capturing the complex relationship between images and metadata, along with the objectives introduced for training.

\subsection{Patient-level Data Pairs}
We construct patient-level data by grouping CFPs for each individual, designing each pair to consist of two images from the same patient, potentially spanning both left and right eyes and different scanners.
When pairs include left and right eye images, we leverage the natural symmetry and biological correlation to extract intrinsic retinal characteristics specific to the individual~\cite{Glaucoma-symmetry,DR-symmetry,AMD-symmetry}. To account for real-world imaging variability, we incorporate images from multiple scanners rather than restricting pairs to a single device, ensuring that the learned representations generalize across diverse acquisition settings. To maximize the number of pairs, we generate all possible combinations of a patient’s retinal images, capturing variations in eye laterality ($l$: left, $r$: right) and scanner type $\mathcal{S}$.
For each patient $i$, we define the image set as $\mathcal{P}_i = \{{I}_{e,s}^{(i)} | e \in \{l, r\}, s \in \mathcal{S}\}$, where ${I}_{e, s}^{(i)}$ represents a $e$-side CFP of patient $i$ acquired by scanner $s$ from the scanner set $\mathcal{S}$.
Each pair is then formed as $({I}_{e_1, s_1}^{(i)}, {I}_{e_2, s_2}^{(i)}$) for some $e_1, e_2 \in \{l, r\}$ and $s_1, s_2 \in \mathcal{S}$, including cases where $s_1 = s_2$ if multiple images from the same scanner are available.
This approach integrates both patient-specific retinal information and inter-scanner variability for effective model training.

\subsection{Retina-Aware Adaptive Masking Strategy}
To account for the distinct properties of retinal images, where the key features primarily reside within the retinal region, we propose a Retina-Aware Adaptive Masking (RAAM). This adaptive masking strategy selectively applies masking to the retinal region and dynamically adjusts the masking ratio during training.

\noindent\textbf{(1) Region-Specific Retinal Masking.}
Traditional MAE training~\cite{mae} applies random masking across the entire image. However, in retinal images, the non-retinal regions (e.g., background) do not contain meaningful information for learning. To ensure that the model focuses on critical retinal structures within the circular retinal image area, we introduce a Region-Specific Masking approach, where the unmasked patches are selectively chosen from within the retinal area (a red circle in Fig.~\ref{fig:arch}).
To implement this, we generate a retinal region mask using the automated retinal image analysis tool, Automorph~\cite{automorph}.

\noindent\textbf{(2) Cosine Decay-Based Masking Ratio Scheduling.}
In retinal image analysis, capturing both global structural patterns and fine-grained details is essential. To achieve this, we introduce a dynamic masking schedule that progressively reduces the masking ratio during training. Starting with a high masking ratio $r_{0}$,  the model is encouraged to prioritize learning broad structural features, minimizing early reliance on localized information~\cite{masking_ratio}. As training progresses, the masking ratio decreases towards the final masking ratio $r_{T}$, allowing the model to focus on fine-grained retinal structures while maintaining strong global representations. 
To ensure stable learning and a smooth transition, we adopt a Cosine Decay-Based Scheduling~\cite{cosine_decay}, which keeps the masking ratio stable early on and reduces it more rapidly as training progresses. 
The masking ratio $r_{t}$ follows a cosine decay function: $r_{t} = 0.5 \left(1 - \cos \left( {\pi t}/{T} \right) \right) (r_{T} - r_{0}) + r_{0}$,
where $T$ is the total number of training epochs.

\subsection{Learnable Metadata Embedding}
To dynamically integrate metadata into the learning process, we introduce the Learnable Metadata Embedding (LME), which enables metadata to adapt and evolve alongside visual features.
We utilize the patient’s age and gender as the sole metadata, keeping the approach lightweight and focused, and introduce learnable meta embeddings, ${m}_{\text{age}}$ and ${m}_{\text{gender}}$, for each attribute.
Formally, we define a set of learnable embeddings $m \in \{{{m}_{\text{age}}, {m}_{\text{gender}}}\}$ and prepend them to the image patch embeddings, ensuring seamless integration with visual features.
After the forward pass, the model produces an output embedding 
for each prompt, which is optimized to align with its corresponding ground-truth metadata labels ($y_{\text{age}}$ or $y_{\text{gender}})$.
Concretely, we optimize these embeddings through a meta loss function $\mathcal{L}_{\text{meta}}$, as outlined in Eq.~\eqref{eq:meta_loss} of the following section. 
By minimizing $\mathcal{L}_{\text{meta}}$, each prompt specializes in representing its corresponding clinical attribute: ${m}_{\text{age}}$ adapts to capture age-related variations, and ${m}_{\text{gender}}$ learns to reflect gender-specific patterns in CFPs.

\subsection{Joint Optimization Objectives}
To train our foundation model, we combine multiple objectives to capture both visual features and metadata. These include reconstruction loss for robust feature extraction, consistency loss to maintain stable representations across patient variations, and meta loss to guide the learnable embeddings for age and gender. 

\noindent\textbf{Reconstruction Loss.}
We include a reconstruction loss, following the objective used in Masked Autoencoders~\cite{mae}, to guide the encoder toward learning discriminative visual features.
Given the necessity of capturing fine-grained anatomical structures in retinal images, we further incorporate perceptual loss \cite{perceptual} to make the model restore subtle details more effectively.

\noindent\textbf{Consistency Loss.}
To further ensure stable learning and robust integration across patient-level variations, we use a feature consistency loss based on cosine similarity: $\mathcal{L}_{\text{consis}} = \frac{1}{N} \sum_{n=1}^{N} \left( 1 - \hat{{z}}_{e_1,s_1}^n \cdot \hat{{z}}_{e_2,s_2}^n \right)$,
where $\hat{{z}}_{e_1,s_1}^n$ are the normalized feature embeddings of an unmasked patch and $\hat{{z}}_{e_2,s_2}^n$ corresponds to the embedding of its paired counterpart in the opposite spatial location. Here, $N$ denotes the total number of unmasked patch pairs across all images in a batch. 

\noindent\textbf{Meta Loss.}
To incorporate clinically relevant metadata into the learning process, we introduce a meta loss that ensures the model effectively captures both continuous and categorical patient attributes:
\begin{equation}
\label{eq:meta_loss}
    \mathcal{L}_{\text{meta}} = \text{RMSE}(\hat{y}_{\text{age}}, y_{\text{age}}) + \text{CE}(\hat{y}_{\text{gender}}, y_{\text{gender}}),
\end{equation}
where $\text{RMSE}$ is the root-mean-square error, applied to enforce accurate age prediction and $\text{CE}$ is the cross-entropy loss used for gender classification.

All loss terms are combined into a weighted sum, where $\lambda_{\text{recon}}$, $\lambda_{\text{consis}}$, $\lambda_{\text{meta}}$ are fixed scaling factors: $\mathcal{L} = \lambda_{\text{recon}}\mathcal{L}_{\text{recon}} + \lambda_{\text{consis}}\mathcal{L}_{\text{consis}} + \lambda_{\text{meta}}\mathcal{L}_{\text{meta}}$.

%% file: sec/3_experiment.tex
\section{Experiments}
We train our foundation model \ourmodel{} on a large-scale dataset and then evaluate it through downstream classification tasks, with minimal additional training.
\input{table/inhouse_table}

\noindent\textbf{Datasets.}
\sloppy
\ourmodel{} was trained on an in-house dataset comprising 1,017,549 CFPs from 292,006 patients across six medical institutions, including UK Biobank. For downstream tasks, we evaluate disease prediction for Diabetic Retinopathy (DR), Glaucoma, and Age-Related Macular Degeneration (AMD), as well as biomarker prediction such as Coronary Artery Calcium (CAC) score and estimated Glomerular Filtration Rate (eGFR) using an in-house dataset. These downstream datasets contain 1,110 to 6,861 CFPs per disease or biomarker, separate from pre-training data, with a 70:15:15 train:val:test split.
Additionally, we conduct disease prediction tasks using public datasets (IDRID \cite{IDRiD}, APTOS2019 \cite{APTOS}, MESSIDOR-2 \cite{MESSIDOR}, GF \cite{GF}, PAPILA \cite{PAPILA}, Retina \cite{Retina}, and JSIEC \cite{JSIEC}).

\noindent\textbf{Training Details.}
The images are resized to $224 \times 224$ and augmented using random crop and flip, with different augmentation ratios applied to each eye. In our pre-training setup, we use a Vision Transformer (ViT) architecture~\cite{vit}, with ViT-S or ViT-B for the encoder and ViT-S for the decoder. The encoder is initialized with ImageNet-1K pre-trained weights~\cite{deng2009imagenet}. 
We train with a batch size of 512 per GPU on 8 H100 GPUs for 300 epochs (15 warm-up). The learning rate is 5e-5. We set the loss function weights as $\lambda_{\text{recon}} = 1.4$, $\lambda_{\text{meta}} = 0.2$, and $\lambda_{\text{consis}} = 0.4$ by default. In reconstruction loss, the original MAE loss weight is 1.0, and the perceptual loss weight is 0.4. We also use RAAM with making ratios of $r_0 = 0.985$ and $r_T=0.85$. Downstream tasks with our pre-trained encoder run for 50 epochs (batch size 16), saving the best validation model for testing.

\subsection{Evaluation Results}

We assess \ourmodel{} against state-of-the-art retinal foundation models, using in-house data (disease/biomarker prediction) and benchmarks (disease prediction).

\input{table/main_table}

\noindent\textbf{Disease/Biomark Prediction.}
We evaluate \ourmodel{} and baselines on internal data for disease and biomarker prediction (Table~\ref{tab:inhouse}). 
For fair comparison based on training data, we also train RETFound on the same data as \ourmodel{} and evaluate its performance.

As a result, \ourmodel{} outperforms baselines in disease prediction tasks and maintains strong performance regardless of encoder size (ViT-S, B). Notably, it surpasses existing foundation models in biomarker prediction tasks, demonstrating its versatility. By effectively integrating metadata, \ourmodel{} leverages patient-specific factors, ensuring robustness across various tasks.

\noindent\textbf{Benchmarks.}
\ourmodel{} is evaluated on disease classification benchmarks consisting of seven public datasets.
Unlike report-supervision models such as FLAIR \cite{flair}, UrFound \cite{urfound}, and RET-CLIP \cite{retclip}, which are trained using clinical descriptions, our model uses only images and metadata. 
Despite this, \ourmodel{} performs comparably to or outperforms baselines, demonstrating generalizability across diverse datasets. These evaluations shows the potential of our foundation models using only metadata, even when clinical text data is unavailable.

\subsection{Ablation Study}
\noindent\textbf{Effect of LME \& RAAM.}
We conduct our experiments by adding RAAM and LME to the SiamMAE on public datasets (Table~\ref{tab:performance_comparison}).
In most datasets, LME and RAAM each strictly improved AUROC and AUPRC, showing significant effects on MESSIDOR-2, PAPILA, Retina, and JSIEC datasets.
Notably, \ourmodel{}, which incorporates both RAAM and LME, achieves the highest performance due to retinal area awareness and the utilization of metadata.

\begin{figure}[t]
  \centering
  \includegraphics[width=\linewidth]{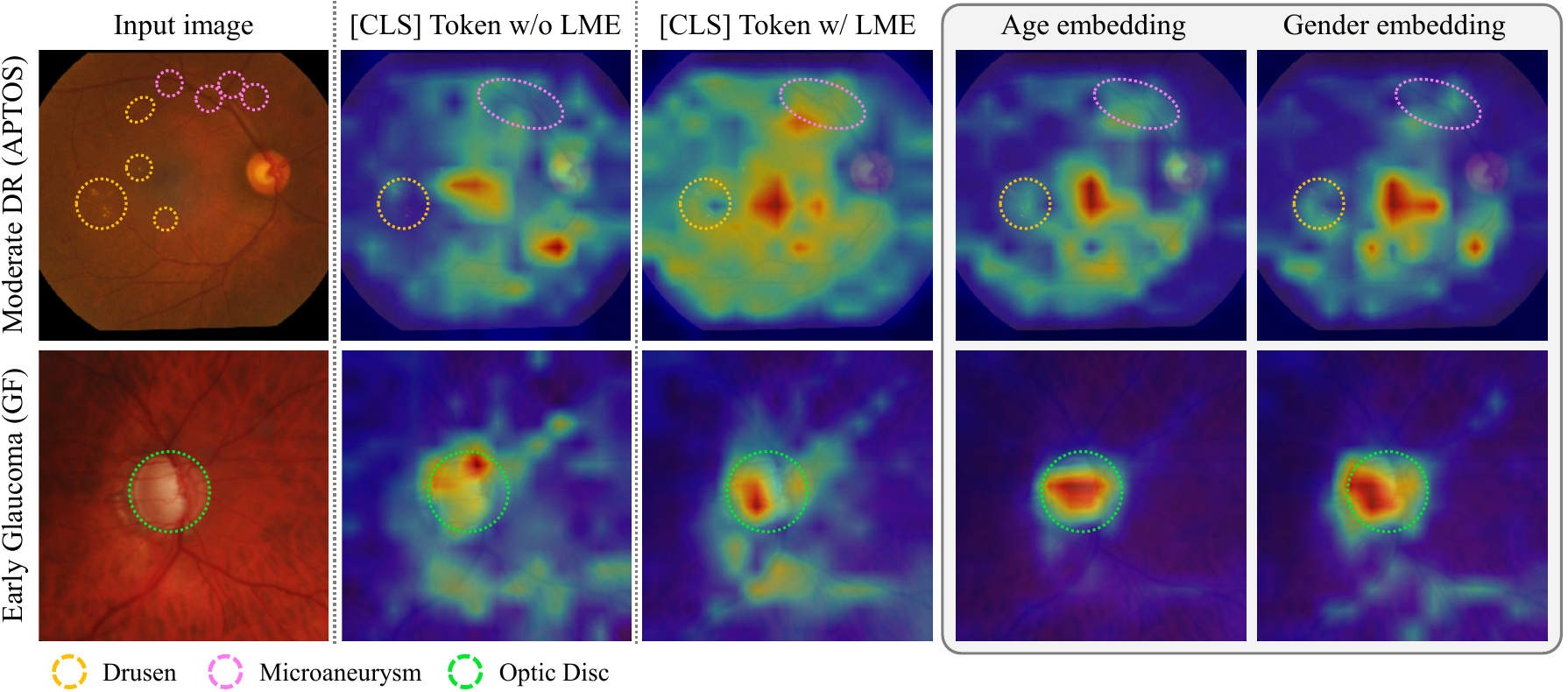}
  \caption{
    Visualization of attention maps for the [CLS] token and metadata embeddings, comparing focus with and without LME.
    These maps highlight regions key to diagnosing eye diseases like diabetic retinopathy and glaucoma.
  }
  \label{fig:attn}
\end{figure}

\noindent\textbf{Impact of LME through Attention Maps.}
To examine how \ourmodel{} utilizes metadata visually, we analyze attention maps. Specifically, we extract and visualize the attention maps of age and gender tokens from \ourmodel{}, along with the CLS token attention maps from both the baseline model without LME and \ourmodel{} with LME, using the APTOS2019 and GF datasets (Fig.~\ref{fig:attn}). In the first row, featuring a retinal image of a patient with moderate diabetic retinopathy (DR), the attention maps of the age and gender tokens in the LME-enabled model precisely highlight structures and features critical for DR diagnosis.
Notably, these metadata attention maps capture not only the macula and vessels but also drusen (yellow circles) and microaneurysms (pink circles), which the model without LME fails to detect but is important for DR~\cite{ablation_dr}. This demonstrates how metadata helps the model effectively recognize key structures essential for DR prediction. Similarly, in the second row, for early glaucoma, age and gender embeddings of the LME-enabled model sharply focus on the optic disc (green circles) and the optic cup within it, both of which are critical for glaucoma diagnosis~\cite{ablation_glaucoma}. Furthermore, in both cases, the CLS token in the LME-enabled model, influenced by metadata embeddings, captures more accurate disease-relevant information for DR and glaucoma, demonstrating that metadata enhances the model’s ability to target critical diagnostic features.

\input{table/ablation_table}

%% file: table/inhouse_table.tex
\begin{table*}[t!]
\caption{Comparison of various methods on disease prediction and biomarker prediction on an in-house dataset. The best AUROC (ROC) and AUPRC (PRC) scores are in \textbf{bold}, while the second-best scores are \underline{underlined}. Our proposed method is highlighted in a pastel color.}
\label{tab:inhouse}
\centering
\setlength{\tabcolsep}{3pt}
\renewcommand{\arraystretch}{1.2}
\resizebox{\textwidth}{!}{
    \begin{tabular}{lcccccccccc}
    \specialrule{0.8pt}{0pt}{0pt}
    \multirow{3}{*}{Method} 
    & \multicolumn{6}{c}{Disease Prediction} 
    & \multicolumn{4}{c}{Biomarker Prediction} \\
    \cline{2-11}
     & \multicolumn{2}{c}{DR} & \multicolumn{2}{c}{Glaucoma} & \multicolumn{2}{c}{AMD} & \multicolumn{2}{c}{CAC} & \multicolumn{2}{c}{eGFR} \\
    \cline{2-11}
     & ROC & PRC & ROC & PRC & ROC & PRC & ROC & PRC & ROC & PRC \\
    \hline
    \rowcolor{gray!20}
    \multicolumn{11}{l}{\textit{train/test on in-house}} \\
    RETFound {\scriptsize (ViT-S)} & 0.871 & 0.586 & 0.531 & 0.508 & 0.488 & 0.499 & 0.783 & 0.623 & 0.614 & 0.598 \\
    \rowcolor{yellow!20} \ourmodel{} {\scriptsize (ViT-S)} & \textbf{0.977} & \textbf{0.866} & \textbf{0.717} & \textbf{0.582} & \textbf{0.751} & \textbf{0.653} & \textbf{0.898} & \textbf{0.732} & \textbf{0.687} & \textbf{0.642}  \\
    \rowcolor{gray!20}
    \multicolumn{11}{l}{\textit{test on in-house}} \\
    MAE {\scriptsize (ViT-B)}~\cite{mae} & 0.965 & 0.818 & 0.697 & 0.590 & \underline{0.729} & 0.616 & 0.805 & 0.620 & 0.504 & 0.497  \\
    RETFound {\scriptsize (ViT-L)}~\cite{retfound} & 0.969 & 0.838 & \underline{0.747} & 0.587 & 0.695 & 0.608 & 0.890 & 0.696 & 0.582 & 0.566   \\
    UrFound {\scriptsize (ViT-B)}~\cite{urfound} & \underline{0.971} & \textbf{0.910} & 0.691 & \underline{0.591} & 0.700 & \underline{0.625} & 0.860 & 0.689 & 0.572 & 0.556   \\
    RET-CLIP {\scriptsize (ViT-B)}~\cite{retclip} & 0.857 & 0.579 & 0.642 & 0.547 & 0.516 & 0.504 & 0.795 & 0.630 & 0.522 & 0.543  \\
    \rowcolor{yellow!20} \ourmodel{} {\scriptsize (ViT-B)} & \textbf{0.982} & \underline{0.885} & \textbf{0.777} & \textbf{0.622} & \textbf{0.772} & \textbf{0.659} & \textbf{0.899}  & \textbf{0.734} & \textbf{0.699} & \textbf{0.643} \\
    \specialrule{0.8pt}{0pt}{0pt}
    \end{tabular}
}
\end{table*}

%% file: table/main_table.tex
\begin{table*}[t!]
\caption{Performance comparison across different datasets and foundation models on public datasets.}
\label{tab:public}
\centering
\setlength{\tabcolsep}{2.5pt}
\renewcommand{\arraystretch}{1.2}
\resizebox{\textwidth}{!}{
    \begin{tabular}{lcccccccccccc}
\specialrule{0.8pt}{0pt}{0pt}
    \multirow{3}{*}{Method}
    & \multicolumn{4}{c}{DR} 
    & \multicolumn{4}{c}{Glaucoma}
    & \multicolumn{4}{c}{Multicategory} \\ 
    \cline{2-13}
    & \multicolumn{2}{c}{APTOS2019} 
    & \multicolumn{2}{c}{MESSIDOR-2} 
    & \multicolumn{2}{c}{GF}
    & \multicolumn{2}{c}{PAPILA}
    & \multicolumn{2}{c}{Retina}
    & \multicolumn{2}{c}{JSIEC} \\
    \cline{2-13}
      & ROC & PRC & ROC & PRC & ROC & PRC & ROC & PRC & ROC & PRC & ROC & PRC \\
    \hline
    \rowcolor{gray!20}
    \multicolumn{13}{l}{\textit{Report Supervision}} \\
    FLAIR~\cite{flair}  & 0.932 & 0.686 & 0.819 & 0.483 & 0.905 & 0.792 & 0.752 & 0.610 & 0.863 & 0.679 & 0.917 & 0.704 \\
    UrFound~\cite{urfound} & 0.949 & 0.716 & 0.882 & 0.608 & 0.958 & 0.880 & 0.783 & 0.625 & 0.901 & 0.793 & 0.995 & 0.923  \\
    RET-CLIP~\cite{retclip} & 0.951 & 0.748 & - & - & 0.958 & 0.889 & 0.853 & 0.754 & 0.942 & 0.871 & 0.999 & 0.972 \\
    \rowcolor{gray!20}
    \multicolumn{13}{l}{\textit{Non-report Supervision}} \\
    MAE~\cite{mae} & 0.941 & 0.676 & 0.842 & 0.488 & 0.931 & 0.832 & 0.629 & 0.475 & 0.742 & 0.537 & 0.985 & 0.818  \\
    RETFound~\cite{retfound} & 0.943 & 0.726 & 0.864 & 0.568 & 0.943 & 0.863 & 0.855 & 0.748 & 0.847 & 0.697 & 0.990 & 0.884 \\
    \rowcolor{yellow!20} \ourmodel{} {\scriptsize (ViT-S)} & \textbf{0.948}  &  0.725  & 0.862  &  0.600 & \textbf{0.955}  &  \textbf{0.888} & 0.838 & 0.714 & 0.831 & 0.656 & 0.990 & 0.853 \\ 
    \rowcolor{yellow!20} \ourmodel{} {\scriptsize (ViT-B)} &\textbf{0.948}  & \textbf{0.743}  & \textbf{0.866}  &  \textbf{0.612} &  \textbf{0.955} & 0.886 & \textbf{0.879} & \textbf{0.764} & \textbf{0.893} & \textbf{0.771} & \textbf{0.994} & \textbf{0.912} \\   
\specialrule{0.8pt}{0pt}{0pt}

    \end{tabular}
}
\end{table*}

%% file: table/ablation_table.tex
\begin{table*}[t!]
\setlength{\tabcolsep}{3pt}
\caption{Comparative analysis of LME and RAAM in the our \ourmodel{} model.}
\label{tab:performance_comparison}
\centering
\renewcommand{\arraystretch}{1.2}
\resizebox{\textwidth}{!}{
\begin{tabular}{lccccccccccccc}
\specialrule{1pt}{0pt}{0pt}
\multirow{3}{*}{Method} &
& \multicolumn{4}{c}{DR} 
& \multicolumn{4}{c}{Glaucoma}
& \multicolumn{4}{c}{Multicategory} \\
\cline{3-14}
&
& \multicolumn{2}{c}{APTOS2019} 
& \multicolumn{2}{c}{MESSIDOR-2} 
& \multicolumn{2}{c}{GF}
& \multicolumn{2}{c}{PAPILA}
& \multicolumn{2}{c}{Retina}
& \multicolumn{2}{c}{JSIEC} \\
\cline{3-14}
 &  & ROC & PRC  & ROC & PRC & ROC & PRC & ROC & PRC & ROC & PRC & ROC & PRC  \\
\hline
Baseline &  & 0.935 & 0.692 & 0.846 & 0.495 & 0.940 & 0.844 & 0.813 & 0.690 & 0.832 & 0.675 & 0.976 & 0.788  \\
+ RAAM &  & 0.943 & 0.713 & 0.864 & 0.585 & 0.950 & 0.875 & 0.845 & 0.736 & 0.861 & 0.717 & 0.989 & 0.857  \\
+ LME &  & 0.943 & 0.724 & 0.851 & 0.534 & 0.953 & 0.883 & 0.864 & 0.740 & 0.871 & 0.718 & 0.989 & 0.875 \\
\rowcolor{yellow!20}
\ourmodel{} &  & \textbf{0.948} & \textbf{0.743} & \textbf{0.866} & \textbf{0.612} & \textbf{0.955} & \textbf{0.886} & \textbf{0.879} & \textbf{0.764} & \textbf{0.893} & \textbf{0.771} & \textbf{0.994} & \textbf{0.912} \\
\specialrule{1pt}{0pt}{0pt}
\end{tabular}
}
\end{table*}

%% file: sec/4_conclusion.tex
\section{Conclusion}
Our \ourmodel{}, a metadata-driven retinal foundation model that leverages LME and RAAM, achieved state-of-the-art performance in disease and biomarker prediction tasks.
Ultimately, \ourmodel{} serves as a patient-specific, metadata-aware foundation model, improving the generalizability by learning ocular disease-related features. 
Using metadata demonstrated effectiveness for retinal analysis, suggesting its potential for broader applications.
While our study focused on age and gender, future work could explore additional metadata types and extend our methodology to other medical imaging domains.